\documentclass{kluwer}    

\newdisplay{guess}{Conjecture}

\def\be{\begin{equation}}
\def\ee{\end{equation}}
\def\bea{\begin{eqnarray}}
\def\eea{\end{eqnarray}}

\begin{document}
\begin{article}
\begin{opening}
\title{A class of exact solutions of the Li\'{e}nard type ordinary
non-linear differential equation}
\author{Tiberiu \surname{Harko}\thanks {email: t.harko@ucl.ac.uk}}
\institute{Department of Mathematics, University College London, Gower Street, London
WC1E 6BT, United Kingdom}
\author{Francisco \surname{S. N. Lobo}\thanks {email: flobo@cii.fc.ul.pt}}
\institute{Centro de Astronomia e Astrof\'{\i}sica da Universidade de Lisboa, Campo
Grande, Ed. C8 1749-016 Lisboa, Portugal}
\author{M. K. \surname{Mak}\thanks {email: mkmak@vtc.edu.hk}}
\institute{Department of Computing and Information Management, Hong Kong Institute of
Vocational Education, Chai Wan, Hong Kong, P. R. China}
\runningauthor{T. Harko, F. S. N. Lobo, M. K. Mak}
\runningtitle{Exact solutions of the Li\'{e}nard type differential equation}
\date{\today}

\begin{abstract}
A class of exact solutions is obtained for the Li\'{e}nard type ordinary
non-linear differential equation. As a first step in our study the second
order Li\'{e}nard type equation is transformed into a second kind Abel type
first order differential equation. With the use of an exact integrability
condition for the Abel equation (the Chiellini lemma), the exact general
solution of the Abel equation can be obtained, thus leading to a class of
exact solutions of the Li\'{e}nard equation, expressed in a parametric form. We also extend the Chiellini integrability condition  to the case of the general Abel equation. As an application of the integrability condition the exact solutions of some
particular Li\'{e}nard type equations, including a generalized van der Pol
type equation, are explicitly obtained.
\end{abstract}

\end{opening}

\section{Introduction}

The Li\'{e}nard type second order nonlinear differential equation of the
form \cite{Lien1, Lien2}
\begin{equation}  \label{1}
\ddot{x}(t)+f(x)\dot{x}(t)+g(x)=0,
\end{equation}
where $f(x)$ and $g(x)$ are arbitrary real functions of $x$, with $%
f(x),g(x)\in C^{\infty }(I)$, defined on a real interval $I\subseteq \Re $,
as well as its generalization, the Levinson-Smith type equation \cite{Lev}
\begin{equation}  \label{lev}
\ddot{x}(t)+f\left(x,\dot{x}\right)\dot{x}(t)+g(x)=0,
\end{equation}
where a dot represents the derivative with respect to the time $t$, and $f$
is a function of $x$ and $\dot{x}$, plays an important role in many areas of electronics (where Eq.~(\ref{1}) appears as the Raylegh
or van der Pol equation), cardiology (modeling the electric heart activity), neurology
(modeling neurons activity), biology, mechanics, seismology, chemistry, physics and cosmology \cite{vdp,2,3,Polh,F,N,Glass,4,5,Pol1,Sal1,Sal2}.

A particular type of the general Li\'{e}nard equation, the van der Pol
equation \cite{vdp,2,3}
\begin{equation}  \label{vdp}
\ddot{x}(t)-\mu \left[1-x^2(t)\right]\dot{x}(t)+x(t)=0,
\end{equation}
where $\mu $ is a positive parameter, describing a non-conservative oscillator with non-linear damping, is
extensively applied in both the physical and biological sciences. In the 1920's the Dutch physicist van der Pol, when he was an engineer working
for Philips Company, studied the differential equation (\ref{vdp}),
which describes the circuit of a vacuum tube. A few years after, the electric activity of the heart rate was modeled by using a Li\'{e}nard type model \cite{Polh}. In the 1960's Fitzhugh \cite{F} and Nagumo et al. \cite{N} extended the van der Pol equation in a planar field as a model for action potentials of neurons. The van der Pol equation, originally introduced to describe relaxation oscillators in electronic circuits, has been frequently
used in theoretical models of the heart function \cite{Glass,4,5}. The van der Pol equation is a useful phenomenological model for the heartbeat,
since it displays many of those features supposedly occurring in the biological setting, as complex periodicity, entrainment, and chaotic behavior \cite{Glass,4,5}.

Most models based on chemical kinetics can be formulated as Li\'{e}nard type coupled nonlinear first-order rate equations in several variables \cite{Pol1}. The first-order approximation for the Li\'{e}nard system works well near a bifurcation point, with higher-order terms being required the further the system is from the bifurcation point. The dynamics of a scalar inflaton field with a symmetric double--well potential can also be formulated mathematically as a Li\'{e}nard system \cite{Sal1,Sal2}. For this case one can  prove rigorously the existence of a limit cycle in its phase space, and, by using analytical and numerical arguments one can show that the limit cycle is stable, and its period can be obtained by an analytical formula.

 The Li\'{e}%
nard type equations can also be used to model fluid mechanical phenomena.
The linearly forced isotropic turbulence can be described in terms of a
cubic Li\'{e}nard equation with linear damping of the form \cite{Ran}
\begin{equation}  \label{dum}
\ddot{x}(t)+\left[ax(t)+b\right]\dot{x}(t)+cx(t)-x^3(t)+d=0,
\end{equation}
where $a$, $b$, $c$ and $d$ are constants, also naturally appear in the mathematical
description of some important astrophysical phenomena. For example, the
time-dependence $\phi (t)$ of the perturbations of the stationary solutions
of spherically symmetric accretion processes can be described by a
generalized Li\'{e}nard type equation of the form \cite{accr}
\begin{equation}
\ddot{\phi }+\epsilon f\left(\phi, \dot{\phi}\right)\dot{\phi }%
+V^{\prime}(\phi )=0,
\end{equation}
where $\epsilon $ is a small parameter, and $V(\phi )$ is the potential of
the system, with the prime indicating the derivative with respect to $\phi $%
. A dynamical systems analysis of this Li\'{e}nard equation reveals a saddle
point in real time, with the implication that when the perturbation is
extended into the nonlinear regime, instabilities will develop in the
accreting system.

From a physical point of view, the Li\'{e}nard equation represents the
generalization of the equation of damped oscillations, $\ddot{x}+\gamma \dot{%
x}+\omega ^2 x=0$, where $\gamma$ and $\omega ^2$ are constant parameters,
respectively \cite{Ben}. For $\gamma =0$ we obtain the equation of the
linear harmonic oscillator, which represents one of the fundamental
equations of both classical and quantum physics. Generally, a linear
oscillation can be described by the equation $\ddot{x}+f(t)\dot{x}+g(t)x=0$.

The mathematical properties of
the Li\'{e}nard type of equations have been intensively investigated from both
mathematical and physical points of view, and their study remains an active
field of research in mathematical physics \cite%
{litm2,litm1,litn,lit0,lit1,lit2,lit3, litp, lit4,lit5,lit6}. Several methods of integrability, like the Lie symmetries method \cite{Lie1,Lie2} and the Weierstrass integrability, introduced in \cite{Wei1}, were used to study the Li\'{e}nard equation, and the relations between the Riccati and Li\'{e}nard equations, respectively.   Li\'enard systems which have a generalized Weierstrass first integral, or a generalized Weierstrass inverse integrating factor, were studied in \cite{Wei2} .

It is the purpose of the present paper to introduce some exactly integrable
classes of the Li\'{e}nard equation, Eq.~(\ref{1}), whose solutions can be
obtained in an exact analytical form, and to formulate the integrability
condition for this class of differential equations. To obtain the functional
form of the integrable Li\'{e}nard type equations we reduce them first to an
Abel type equation of the form $y'=p(x)y^3+q(x)y^2$ \cite{Pol, kamke}. Then, we apply to the latter Abel equation an
integrability condition, equivalent to the initial Li\'{e}nard equation,
that was obtained by Chiellini \cite{Chiel, kamke}. In fact, the Chiellini
condition has been recently used for the study of the Abel differential
equations, as well as for the second order differential equations reducible
to an Abel type equation, in \cite{Ban, Mak1,Mak2,Mak3, Rosu1,Rosu2}.

Bandic \cite{Ban} studied the non-linear differential equation $y''+\psi (y)y'^2+\phi (y)y'+f(y)=0$, and he did show that it can be solved by using quadratures if $f(y)=F(y)\exp \left(-2\int{\psi (y)dy}\right)$ and $\phi (y)=G(y)\exp\left(-\int{\psi (y)dy}\right)$, where $F(y)$ and $G(y)$ are the coefficients of the integrable Abel equation $w'=F(y)w^3+G(y)w^2$. The integrability of this Li\'{e}nard type equation was obtained by using the Chiellini condition.

In \cite{Rosu1} the differential Chiellini integrability condition was reformulated in an integral form, and the general form of the solution of the Abel equation was obtained in a simpler form. The Chiellini integrability condition of the first order first kind Abel equation $dy/dx=f(x)y^2+g(x)y^3$ was extended to the case of the general Abel equation of the form $dy/dx=a(x)+b(x)y+f(x)y^{\alpha -1}+g(x)y^{\alpha }$, where $\alpha \in \Re $ and $\alpha >1$, in \cite{Har2}.

There are several other methods that can be used for the integration of the general Abel type equation $y'=p(x)y^3 + q(x)y^2 + r(x)y + s(x)$ \cite{kamke}. If $y = y_l(x)$ is a particular solution of the general Abel equation, then by means of the
transformations $u(x) = E(x)/\left[y(x)-y_1(x)\right]$, where $E(x) = \exp \{\int{\left[3p(x)y_1 ^2 + 2q(x)y_1 + r(x)\right] dx} \}$, the Abel can be transformed into $du/dx+\Phi _1/u+\Phi _2=0$, where $\Phi _1 x)=p(x)E^2(x)$, and $\Phi _2(x) = \left[3p(x)y_(x) + q(x)\right]E(x)$. Therefore, if
$y_l=-q(x)/ 3p(x)$,
then $\Phi_2 = 0$, and the general solution of the Abel equation can be obtained from the integration of a
differential equation with separable variables.

Therefore it turns
out that, if the coefficients $f(x)$ and $g(x)$ of the Li\'{e}nard equation
satisfy two specific conditions, then the general solution of the Li\'{e}%
nard equation can be obtained in an exact analytical form. Some examples of
exactly integrable Li\'{e}nard equations, of physical interest, are also
considered. The generalization of the method to the case of the
Levinson-Smith type equations of the form (\ref{lev}) is briefly discussed.

The present paper is organized as follows. In Section~\ref{sect2}, we
introduce the Abel equation representation for the Li\'{e}nard equation, and
we formulate the integrability condition of the first order Abel equation.
In Section~\ref{sect3}, we obtain the general solution of the Li\'{e}nard
type equations satisfying the integrability condition of the Abel equation.
The exact solutions of some non-linear Li\'{e}nard type differential
equations are obtained in Section~\ref{sect4}. We discuss and conclude our
results in Section~\ref{sect5}.

\section{Reduction of the Li\'{e}nard equation to an integrable Abel type
equation}
\label{sect2}

As a first step in our study of the Li\'{e}nard equation (\ref{1}) we reduce
it to an Abel type first order non-linear differential equation. Then, by
using an integrability condition for this equation, which involves a
differential relation between the coefficients $f(x)$ and $g(x)$ of the
equations, we obtain the general solution of the Li\'{e}nard equation in an
exact parametric form.

By denoting $\dot{x}=u$, the Li\'{e}nard equation (\ref{1}) can be written
as
\begin{equation}  \label{2}
u\frac{du}{dx}+f(x)u+g(x)=0.
\end{equation}
By introducing a new dependent variable $v=1/u$, Eq.~(\ref{2}) takes the
form of the standard first kind Abel differential equation,
\begin{equation}  \label{3}
\frac{dv}{dx}=f(x)v^2+g(x)v^3.
\end{equation}

\subsection{The Chiellini integrability condition for the reduced Abel equation}

In this context, an exact integrability condition for the Abel equation Eq.~(%
\ref{3}) was obtained by Chiellini \cite{Chiel} (see also \cite{kamke}), and
can be formulated as the Chiellini Lemma as follows:

\textbf{Chiellini Lemma}. If the coefficients $f(x)$ and $g(x)$ of a first
kind Abel type differential equation of the form
\begin{equation}  \label{4}
\frac{dv}{dx}=f(x)v^2+g(x)v^3,
\end{equation}
satisfy the condition
\begin{equation}  \label{5}
\frac{d}{dx}\frac{g(x)}{f(x)}=kf(x),
\end{equation}
where $k=\mathrm{constant}\neq 0$, then the Abel Eq.~(\ref{4}) can be exactly integrated.

In order to prove the Chiellini Lemma we introduce a new dependent variable $%
w$ defined as \cite{Chiel, kamke}
\begin{equation}
v=\frac{f(x)}{g(x)}w.
\end{equation}

Then Eq.~(\ref{4}) can be written as
\begin{equation}
\left[ \frac{1}{g(x)}\frac{df(x)}{dx}-\frac{f(x)}{g^{2}(x)}\frac{dg(x)}{dx}%
\right] w+\frac{f(x)}{g(x)}\frac{dw}{dx}=\frac{f^{3}(x)}{g^{2}(x)}\left(
w^{3}+w^{2}\right) .  \label{6}
\end{equation}%
On the other hand, the condition given by Eq.~(\ref{5}) can be written in an
equivalent form as
\begin{equation}
\frac{f(x)}{g^{2}(x)}\frac{dg(x)}{dx}-\frac{1}{g(x)}\frac{df(x)}{dx}=k\frac{%
f^{3}(x)}{g^{2}(x)}.
\end{equation}%
Therefore Eq.~(\ref{6}) becomes
\begin{equation}
\frac{dw}{dx}=\frac{f^{2}(x)}{g(x)}w\left( w^{2}+w+k\right) ,  \label{6_1}
\end{equation}%
which is a first order separable differential equation, with the general
solution given by
\begin{equation}
\int {\frac{f^{2}(x)}{g(x)}dx}=\int {\frac{dw}{w\left( w^{2}+w+k\right) }}\equiv F(w,k).
\label{7}
\end{equation}

With the use of the condition (\ref{5}), the left hand side of Eq.~(\ref{7})
can be written as \cite{Rosu1}
\begin{equation}
\int {\frac{f^{2}(x)}{g(x)}dx}=\frac{1}{k}\int {\frac{d}{dx}\ln }\left|
\frac{g(x)}{f(x)}\right| {dx}=\frac{1}{k}\ln \left| \frac{g(x)}{f(x)}\right|
+C_{0},
\end{equation}%
where $C_{0}$ is an arbitrary constant of integration. Therefore the general
solution of Eq.~(\ref{6_1}) is obtained as
\begin{equation}
\frac{g(x)}{f(x)}=C^{-1}e^{F(w,k)},  \label{sol}
\end{equation}%
where $C^{-1}=\exp \left(- kC_{0}\right) $ is an arbitrary constant of
integration, and
\begin{equation}
e^{F(w,k)}=\left\{
\begin{array}{lll}
\frac{w}{\sqrt{w^{2}+w+k}}\exp \left( -\frac{1}{\sqrt{4k-1}}\arctan \frac{%
1+2w}{\sqrt{4k-1}}\right) , k>\frac{1}{4}, & & \\
&  &  \\
\exp \left[ \frac{1}{1+2w}-2\mathrm{arctanh}(1+4w)\right], k=\frac{1}{%
4}, \label{sol2} &  &  \\
&  &  \\
\frac{w}{\sqrt{w^{2}+w+k}}\left( 1-\frac{1+2w}{\sqrt{1-4k}}\right) ^{-1/2%
\sqrt{1-4k}}\left( 1+\frac{1+2w}{\sqrt{1-4k}}\right) ^{1/2\sqrt{1-4k}%
}, k<1/4, &  &
\end{array}%
\right.
\end{equation}
respectively. Eq.~(\ref{sol}) determines $w$ as a function of $x$.

The integrability condition given by Eq.~(\ref{5}) can be written as
\begin{equation}
\frac{dg(x)}{dx}=\frac{1}{f(x)}\frac{df(x)}{dx}g(x)+kf^2(x),
\end{equation}
representing a first order linear differential equation in $g$. As a
function of $g$, the function $f$ satisfies the differential equation
\begin{equation}  \label{c2}
\frac{1}{f(x)}\frac{df(x)}{dx}=-k\frac{1}{g(x)}f^2(x)+\frac{1}{g(x)}\frac{%
dg(x)}{dx}.
\end{equation}

In order to solve Eq.~(\ref{c2}) we introduce a new dependent variable $%
f(x)=1/\sigma (x)$, and  by denoting $\sigma ^2(x)=\xi (x)$, we obtain a first order differential
equation for $\xi$,
\begin{equation}
\frac{d\xi (x)}{dx}=-2\frac{1}{g(x)}\frac{dg(x)}{dx}\xi (x)+\frac{2k}{g(x)}.
\end{equation}

Therefore the Chiellini Lemma can be reformulated as:

\textbf{Lemma 1.} If the coefficients $f(x)$ and $g(x)$ of the Abel Eq.~(\ref%
{4}) satisfy the conditions
\begin{equation}
g(x)=f(x)\left[ C_{1}+k\int {f(x)dx}\right] ,
\end{equation}%
or
\begin{equation}
f(x)=\pm \frac{g(x)}{\sqrt{C_{2}+2k\int {g(x)dx}}},
\end{equation}%
where $C_{1}$, $C_{2}$, and $k$ are arbitrary constants, the Abel equation
is exactly integrable, and its solution is given by
\begin{equation}
v(x)=Ce^{-F(w(x),k)}w(x),
\end{equation}%
where the functions $F(w,k)$ are given by Eqs.~(\ref{sol2}). A similar result was obtained in \cite{Rosu1}.

\subsection{The Chiellini integrability condition for the general Abel equation}

The Chiellini Lemma can be extended to the general Abel equation of the form
\begin{equation}
\frac{dv}{dx}=a(x)+b(x)v+f(x)v^{2}+g(x)v^{3},  \label{gen}
\end{equation}%
where $a(x),b(x),f(x),g(x)\in C^{\infty }(I)$ are defined on a real interval
$I\subseteq \Re $, and $a(x),b(x)\neq 0,\forall x\in I$, as follows. By
introducing a new function $p(x)$, defined as
\begin{equation}
v(x)=e^{\int {b(x)dx}}p(x),
\end{equation}%
Eq.~(\ref{gen}) becomes
\begin{equation}
\frac{dp}{dx}=a(x)e^{-\int {b(x)dx}}+f(x)e^{\int {b(x)dx}}p^{2}+g(x)e^{2\int
{b(x)dx}}p^{3}.  \label{gen1}
\end{equation}%
We assume now that the functions $b(x)$, $f(x)$ and $g(x)$ satisfy the
condition
\begin{equation}
\frac{d}{dx}\frac{g(x)e^{\int {b(x)dx}}}{f(x)}=k_{1}f(x)e^{\int {b(x)dx}},
\label{cond1}
\end{equation}%
where $k_{1}$ is an arbitrary constant. Then, by introducing the
transformation
\begin{equation}
p(x)=\frac{f(x)}{g(x)e^{\int {b(x)dx}}}s(x),
\end{equation}%
Eq.~(\ref{gen1}) becomes
\begin{equation}
\frac{ds}{dx}=\frac{g(x)}{f(x)}a(x)+\frac{f^{2}(x)}{g(x)}s\left(
s^{2}+s+k_{1}\right) .
\end{equation}

Hence we have obtained the following generalization of the Chiellini Lemma:

\textbf{Lemma 2.} If the coefficients of the general Abel Eq.~(\ref{gen})
satisfy the conditions (\ref{cond1}) and
\begin{equation}
a(x)=k_{2}\frac{f^{3}(x)}{g^{2}(x)},
\end{equation}%
where $k_{2}$ is an arbitrary constant, then the Abel equation can be
exactly integrated, and its general solution is given by
\begin{equation}
v(x)=\frac{f(x)}{g(x)}s(x),
\end{equation}%
with $s(x)$ a solution of the equation
\begin{equation}
G_{0}(s,k_{1},k_{2})=\int {\frac{f^{2}(x)}{g(x)}dx=}\frac{1}{k_{1}}\ln
\left| \frac{g\left( x\right) e^{\int b\left( x\right) dx}}{f\left( x\right)
}\right| +K_{0},  \label{gg}
\end{equation}%
where $K_{0}$ is an arbitrary constant of integration, and
\begin{equation}
G_{0}\left( s,k_{1},k_{2}\right) =\int {\frac{ds}{s^{3}+s^{2}+k_{1}s+k_{2}}}.
\end{equation}

By using Eq.~(\ref{cond1}), Eq. (\ref{gg}) becomes
\begin{equation}
\frac{g\left( x\right) e^{\int {b(x)dx}}}{f\left( x\right) }
=K_{1}e^{G\left(s,k_{1},k_2 \right)},
\end{equation}
where $K_{1}=\exp \left( - k_{1}K_{0}\right) $ is an arbitrary constant of
integration, and $G\left( s,k_{1},k_{2}\right) =k_{1}G_{0}\left(
s,k_{1},k_{2}\right) $, respectively.

\section{A class of exact solutions of the Li\'{e}nard equation}
\label{sect3}

As we have already seen, the second order non-linear Li\'{e}nard Eq.~(\ref{1}%
) can be reduced to an Abel type equation of the form given by Eq.~(\ref{4}%
), with the general solution given by $v(x)=C\exp \left[ -F(w(x),k)\right]
w(x)$, where $w(x)$ is determined, as a function of $x$, by Eq.~(\ref{sol}).
Alternatively, Eq.~(\ref{sol}) fixes $x$ as a function of $w$,
\begin{equation}
x=x(w).
\end{equation}%
To find the time dependence of $x$, we start from
\begin{equation}
\frac{dx}{dt}=\frac{dx}{dw}\frac{dw}{dt}=u=\frac{1}{v}=\frac{g(x)}{f(x)}%
\frac{1}{w},  \label{29}
\end{equation}%
which gives
\begin{equation}
\frac{dw}{dt}=\frac{dw}{dx}\frac{g(x)}{f(x)}\frac{1}{w}.
\end{equation}%
With the use of Eq.~(\ref{6_1}), satisfied by the function $w(x)$, we obtain
for $dw/dt$ the equivalent expression,
\begin{equation}
\frac{dw}{dt}=f(x)\left( w^{2}+w+k\right) .
\end{equation}%

Therefore we have obtained the following:

\textbf{Theorem}. If the coefficients of the Li\'{e}nard equation (\ref{1})
satisfy the conditions
\begin{equation}
g(x)=f(x)\left[ C_{1}+k\int {f(x)dx}\right] ,  \label{32}
\end{equation}%
or
\begin{equation}
f(x)=\pm \frac{g(x)}{\sqrt{C_{2}+2k\int {g(x)dx}}},  \label{33}
\end{equation}%
where $C_{1}$, $C_{2}$ and $k$ are arbitrary constants, then the general
solution of the Li\'{e}nard equation Eq.~(\ref{1}) can be obtained in an exact parametric
form, with $w$ taken as a parameter, as
\begin{equation}
t-t_{0}=\int {\frac{dw}{f(x(w))\left( w^{2}+w+k\right) }},\qquad x=x(w),
\label{34}
\end{equation}%
with $x(w)$ obtained as a solution of the equation
\begin{equation}
\frac{g(x)}{f(x)}=C^{-1}e^{F(w,k)},  \label{solr}
\end{equation}%
and
\begin{equation}
F(w,k)=k\int {\frac{dw}{w\left( w^{2}+w+k\right) }}.
\end{equation}

Similar results were previously obtained in \cite{Rosu1}, where the integral form of the Chiellini integrability condition were explicitly formulated. A particular integrable case of the Li\'{e}nard equation can be obtained for
the case $k=0$. In this case, the Chiellini condition given by Eq.~(\ref{5})
immediately provides
\begin{equation}
g(x)=Af(x),
\end{equation}
with $A$ an arbitrary constant. Thus, the Li\'{e}nard equation takes the
particular form
\begin{equation}  \label{lies}
\ddot{x}+f(x)\dot{x}+Af(x)=0,
\end{equation}
with the associated Abel equation given by
\begin{equation}  \label{eqs}
\frac{dv}{dx}=f(x)v^2\left(1+Av\right),
\end{equation}
where $\dot{x}=1/v$. The general solution of Eq.~(\ref{eqs}) is given by
\begin{equation}  \label{eqa}
\int{f(x)dx}=A\ln\left|\frac{1}{v}+A\right|-\frac{1}{v}+K_1,
\end{equation}
where $K_1$ is an arbitrary constant of integration. Therefore, the general
solution of Eq.~(\ref{lies}) can be written in a parametric form, with $v$
taken as a parameter, in the following form
\begin{equation}  \label{42}
t-t_0=\int{\frac{dv}{f(x(v))v\left(1+Av\right)}}, \qquad x=x(v),
\end{equation}
where $x=x(v)$ is the solution of Eq.~(\ref{eqa}).

In the general solution for the time, given by Eqs.~(\ref{34}) and (\ref{42}%
), one can take the arbitrary integration constant $t_0$ as zero, without
any loss of generality. This choice fixes the origin of time at $t=0$. The
arbitrary integration constant $C$, as well as the initial value $w_0$ of
the parameter $w$ can be determined from the initial conditions at $t=0$ for
the position $x$ and the velocity $\dot{x}$, given by
\begin{equation}
x(0)=x_0, \qquad \dot{x}(0)=\dot{x}_0,
\end{equation}
where $x_0$ and $\dot{x}_0$ are the initial values of $x$ and $\dot{x}$ at $%
t=0$. By evaluating Eq.~(\ref{solr}) for $x=x_0$, we obtain
\begin{equation}
\frac{g\left(x_0\right)}{f\left(x_0\right)}=C^{-1}e^{F\left(w_0,k\right)},
\end{equation}
while evaluating Eq.~(\ref{29}) at $t=0$ gives the equation $\dot{x}_0=\left[%
g\left(x_0\right)/f\left(x_0\right)\right]w_0^{-1}$, which determines the
initial value of the parameter $w_0$ as
\begin{equation}
w_0=\frac{1}{\dot{x}_0}\frac{g\left(x_0\right)}{f\left(x_0\right)}.
\end{equation}
Once the initial value of the parameter $w_0$ is known, the value of the
integration constant $C^{-1}$ is obtained as
\begin{equation}
C^{-1}=\frac{g\left(x_0\right)}{f\left(x_0\right)}e^{-F\left(w_0,k\right)}.
\end{equation}

\subsection{An integrability condition for the Levinson-Smith equation}

The procedure for the exact integration of the Li\'{e}nard type equations
based on the Chiellini Lemma can be easily extended to the generalized Li%
\'{e}nard equations of the Levinson-Smith form, given by Eq.~(\ref{lev}), if
they can be transformed to an Abel type equation. As a particular case of
the integrable Levinson-Smith type non-linear differential equations we
consider the equation
\begin{equation}  \label{ls}
\ddot{x}+\left[\gamma (x)\dot{x}^2+\delta (x)\dot{x}+f(x)\right]\dot{x}%
+g(x)=0,
\end{equation}
where $\gamma (x)$ and $\delta (x)$ are some arbitrary functions of the
variable $x$. By denoting $\dot{x}=1/v$, Eq.~(\ref{ls}) takes the form of
the general Abel equation
\begin{equation}  \label{ls1}
\frac{dv}{dx}=\gamma (x)+\delta (x)v+f(x)v^2+g(x)v^3=0.
\end{equation}

If $\gamma (x)=0$, by means of the transformation $v(x)=e^{\int {\delta (x)dx%
}}h(x)$, Eq.~(\ref{ls1}) can be written in the standard form of the Abel
equation,
\begin{equation}
\frac{dh}{dx}=A(x)h^{2}+B(x)h^{3},  \label{abells}
\end{equation}%
where $A(x)=f(x)e^{\int {\delta (x)dx}}$, and $B(x)=g(x)e^{2\int {\delta
(x)dx}}$. If the coefficients $A(x)$ and $B(x)$ of the equation satisfy the
conditions of Lemma 1, then the general solution of Eq.~(\ref{abells}) can
be obtained through quadratures. If $\gamma (x)\neq 0$, then from Lemma 2 it
follows that if the functions $\gamma (x)$, $\delta (x)$, $f(x)$ and $g(x)$
satisfy the conditions
\begin{equation}
\frac{d}{dx}\frac{g(x)e^{\int {\delta (x)dx}}}{f(x)}=k_{1}f(x)e^{\int {%
\delta (x)dx}},  \gamma (x)=k_{2}\frac{f^{3}(x)}{g^{2}(x)},
\end{equation}
where $k_1$ and $k_2$ are two arbitrary constants, then the generalized Li\'{e}nard type equation (\ref{ls}) can be integrated
exactly. Therefore all the integrability results obtained for the Li\'{e}%
nard equation can be applied for the Levinson-Smith type equations of the
form (\ref{ls}).

\section{Examples of exactly integrable Li\'{e}nard type equations}

\label{sect4}

In the present Section, we consider some exactly integrable Li\'{e}nard type
equations, which represent the generalizations of Eqs.~(\ref{vdp}) and (\ref%
{dum}), respectively. As a first case we assume that the functional form of
the function $f(x)$ is known. Then the Chiellini integrability condition
fixes the form of the function $g(x)$, and allows to find the general
solution of the Li\'{e}nard equation in an exact parametric form. The case
in which the function $g(x)$ is fixed is also considered. Furthermore, an
integrable generalization of the van der Pol equation is also explored.

\subsection{First case: $f(x)=ax+b$.}

As a first case we assume that the function $f(x)$ is given by
\begin{equation}
f(x)=ax+b,
\end{equation}
where $a$ and $b$ are arbitrary constants.
Then from the first integrability condition, given by Eq.~(\ref{32}) we
obtain the function $g(x)$ as
\begin{equation}
g(x)=\frac{1}{2} a^2 k x^3+\frac{3}{2} a b k x^2+ \left(a C_1+b^2
k\right)x+b C_1,
\end{equation}
where $C_1$ and $k$ are arbitrary integration constants. Therefore the
exactly integrable Li\'{e}nard equation is given by
\begin{equation}  \label{48}
\ddot{x}+\left(ax+b\right)\dot{x}+\frac{1}{2} a^2 k x^3+\frac{3}{2} a b k
x^2+ \left(a C_1+b^2 k\right)x+b C_1=0.
\end{equation}

As a function of the parameter $w$, $x$ is determined by Eq.~(\ref{solr}),
which gives for $x$ the quadratic algebraic equation

\begin{equation}
\frac{ak}{2}x^{2}+bkx+C_{1}=C^{-1}e^{F(w,k)},
\end{equation}%
which determines $x$ as a function of $w$ as
\begin{equation}
x(w)=\frac{-bk\pm \sqrt{b^{2}k^{2}-2ak\left[ C_{1}-C^{-1}e^{F(w,k)}\right] }%
}{ak}.  \label{50}
\end{equation}%
The time dependence of $x$ is determined as a function of $w$ as
\begin{equation}
t-t_{0}=\pm \int {\frac{kdw}{\sqrt{b^{2}k^{2}-2ak\left[
C_{1}-C^{-1}e^{F(w,k)}\right] }\left( w^{2}+w+k\right) }}.  \label{51}
\end{equation}%
Eqs.~(\ref{50}) and (\ref{51}) give the exact solution of the Li\'{e}nard
Eq.~(\ref{48}). Depending on the value of the constant $k$ there are three
distinct classes of solutions of this equation.

\subsection{Second case: $g(x)=cx+d$}

Secondly, we consider the case in which the function $g(x)$ is fixed. By
analogy with the van der Pol Eq.~(\ref{vdp}), we assume that
\begin{equation}
g(x)=cx+d,
\end{equation}%
where $c$ and $d$ are arbitrary constants. Then, after determining the function $f(x)$ from the integrability condition
Eq.~(\ref{33}), we obtain the Li\'{e}nard equation
\begin{equation}
\ddot{x}\pm \frac{cx+d}{\sqrt{ckx^{2}+2dkx+C_{2}}}\dot{x}+cx+d=0.  \label{53}
\end{equation}%
Eq.~(\ref{solr}) gives the equation
\begin{equation}
ckx^{2}+2dkx+C_{2}=C^{-2}e^{2F(w,k)},
\end{equation}%
with the solution
\begin{equation}
x(w)=\frac{-dk\pm \sqrt{d^{2}k^{2}-ck\left[ C_{2}-C^{-2}e^{2F(w,k)}\right] }%
}{ck}.  \label{55}
\end{equation}%
The parametric time dependence of the solution is obtained as
\begin{equation}
t-t_{0}=\pm \frac{k}{C}\int {\frac{e^{F(w,k)}dw}{\sqrt{d^{2}k^{2}-ck\left[
C_{2}-C^{-2}e^{2F(w,k)}\right] }\left( w^{2}+w+k\right) }}.  \label{56}
\end{equation}%
Eqs.~(\ref{55}) and (\ref{56}) give the exact analytic solution, in a
parametric form, of the Li\'{e}nard Eq.~(\ref{53}).

\subsection{Third case: the generalization of the van der Pol equation}

Finally, we consider the integrable generalization of the van der Pol Eq.~(%
\ref{vdp}), in which we fix the function $f(x)$ as $f(x)=-\mu \left(
1-x^{2}\right) $, and obtain the function $g(x)$ from the integrability
condition Eq.~(\ref{32}). Therefore the integrable generalization of the van
der Pol equation is given by
\begin{equation}
\ddot{x}-\mu \left( 1-x^{2}\right) \dot{x}+\frac{1}{3}k\mu ^{2}x^{5}-\frac{4%
}{3}k\mu ^{2}x^{3}+C_{1}\mu x^{2}+k\mu ^{2}x-C_{1}\mu =0.
\end{equation}%
The parametric dependence of $x$ is determined from the algebraic equation
\begin{equation}
C_{1}-k\mu x+\frac{1}{3}k\mu x^{3}=C^{-1}e^{F(w,k)}.  \label{eqc}
\end{equation}

Equation~(\ref{eqc}) can be rewritten in the form
\begin{equation}
x^{3}-3x+H\left( F\right) =0.  \label{alg}
\end{equation}%
where we have denoted
\begin{equation}
H\left( F(w,k)\right) =\frac{3\left[ CC_{1}-e^{F(w,k)}\right] }{Ck\mu }.
\end{equation}%
The solution of the algebraic Eq.~(\ref{alg}) is given by
\bea
x\left( w\right) &=&\frac{2^{1/3}}{\left\{ \sqrt{H^{2}\left( F(w,k)\right) -4}%
-H\left( F(w,k)\right) \right\} ^{1/3}}+\nonumber\\
&&\frac{\left\{ \sqrt{H^{2}\left(
F(w,k)\right) -4}-H\left( F(w,k)\right) \right\} ^{1/3}}{2^{1/3}}.
\label{cubic}
\eea

In order to have a real solution of the cubic Eq.~(\ref{alg}), the
conditions
\begin{equation}
H^{2}\left( F(w,k)\right) -4>0,
\end{equation}%
and
\begin{equation}
\sqrt{H^{2}\left( F(w,k)\right) -4}-H\left( F(w,k)\right) >0,
\end{equation}%
must be satisfied for all $w$ and $k$.

The parametric time dependence of the solution of the generalized van der
Pol equation is obtained as
\begin{equation}
t-t_{0}=\frac{1}{\mu }\int {\frac{\phi ^{2}(w,k)dw}{\left[ \phi
^{4}(w,k)+\phi ^{2}(w,k)+1\right] \left( w^{2}+w+k\right) }},
\end{equation}
where we have denoted
\begin{equation}
\phi \left( w,k\right) =\frac{2^{1/3}}{\left( \sqrt{H^{2}\left( F(w,k)\right)%
-4}-H\left( F(w,k)\right)\right) ^{1/3}}.
\end{equation}

Depending on the numerical values of the parameters $k,\mu ,C_{1},C$ a large
class of dynamical evolutions of the solutions of the generalized van der
Pol equation can be obtained.

\section{Discussions and final remarks}

\label{sect5}

In the present paper we have introduced a class of exactly integrable Li\'{e}%
nard, and generalized Li\'{e}nard type equations. If the coefficients of the
second order non-linear equations satisfy some specific conditions, which
follow from the Chiellini Lemma, then the general solution of the Li\'{e}%
nard differential equation can be obtained in an exact parametric form. As
an application of the integrability procedure obtained, we have considered
some specific examples of exactly integrable non-linear differential
equations that could be of physical interest. One of these equations, Eq.~(%
\ref{48}) is similar in form with Eq.~(\ref{dum}), and in fact represents
the exactly solvable generalization of the equation describing the linearly
forced isotropic turbulence \cite{litm2}. We have also considered an exactly
solvable generalization of the classical van der Pol oscillator equation, in
which higher order force terms are also present. In all these cases of
physical interest the general solution of the corresponding Li\'{e}nard
equation can be obtained in an exact parametric form. The existence of an
analytical solution may allow a deeper understanding of the highly
non-linear physical processes that govern most of the natural phenomena.

The exact solutions also allow us to obtain some approximate solutions of
the considered differential equations, corresponding to the small and large
values of the parameter $w$, respectively. In the limit of small $w$, i.e., $%
w\ll k$, giving $\exp \left[ F(w,k)\right] \approx w$, Eq.~(\ref{solr})
takes the simple form
\begin{equation}
\frac{g(x)}{f(x)}\approx C^{-1}w,
\end{equation}%
while the parametric time evolution can be obtained as
\begin{equation}
t-t_{0}\approx \frac{1}{k}\int {\frac{dw}{f(x(w))}}.
\end{equation}

In the limit of large $w$, so that $w\gg k$, and $w^2 \gg w$, $\exp \left(
F(w,k)\right) \approx \exp\left(k\int{dw/w^3}\right)=\exp \left(
-k/2w^{2}\right) $, and the approximate asymptotic solution of the exactly
integrable Li\'{e}nard equation is given by
\begin{equation}
\frac{g(x)}{f(x)}\approx C^{-1}e^{-k/2w^{2}},
\end{equation}
and
\begin{equation}
t-t_{0}\approx \int {\frac{dw}{f(x(w))w^{2}}}.
\end{equation}

As an application of the previous asymptotic equations we consider the case $%
f(x)=ax+b$, with $a,b=\mathrm{constant}$, giving $g(x)/f(x)=C_1+kax^2/2+kbx$.

In the limit of small $x$, by neglecting the $x^{2}$ term, we obtain
\begin{equation}
x(w)\approx \frac{C^{-1}w-C_{1}}{bk},
\end{equation}%
and
\begin{equation}
t-t_{0}\approx \frac{bC}{a}\ln \left| \frac{aw}{C}-aC_{1}+b^{2}k\right| ,
\end{equation}
respectively, giving
\begin{equation}
x(t)\approx \frac{1}{abk}e^{a\left( t-t_{0}\right) /bC}-\frac{b}{a}.
\end{equation}

In the limit of large $x$, so that $kbx\gg C_{1}$, and $ax/2\gg b$,
respectively, we obtain $g(x)/f(x)\approx kax^{2}/2$, and
\begin{equation}
x(w)\approx \sqrt{\frac{2C^{-1}}{ka}}e^{-k/4w^{2}},
\end{equation}%
\begin{equation}
t-t_{0}\approx \int {\frac{dw}{w^{2}\left( \sqrt{2a/kC}\;e^{-k/4w^{2}}+b%
\right) }}.
\end{equation}%
In the range of the values of $w$ for which $\sqrt{2a/kC}\;e^{-k/4w^{2}}\gg
b $, we obtain
\begin{equation}
t-t_{0}\approx -\sqrt{\frac{\pi C}{2a}}\;\mathrm{erfi}\left( \frac{\sqrt{k}}{%
2w}\right) ,
\end{equation}%
where $\mathrm{erfi}(z)$ gives the imaginary error function $\mathrm{erfi}%
(z)=\mathrm{erf}(iz)/i$. In the large time limit the solution of the Li\'{e}%
nard Eq.~(\ref{48}) can be obtained only in a parametric form.

\acknowledgements
We would like to thank the four anonymous referees for comments and suggestions that helped us to improve our manuscript.


\end{article}
\end{document}